\documentclass{elsart}

 \usepackage{graphicx}

\usepackage{amssymb}
\usepackage{amsmath}

\begin{document}

\begin{frontmatter}



\title{Thresholds, long delays and stability from generalized allosteric effect in protein networks}


\author[Verona,INFN]{Roberto Chignola}
\ead{roberto.chignola@univr.it}
\author[Verona]{Chiara Dalla Pellegrina}
\author[Trieste,INFN]{Alessio Del Fabbro}
\ead{delfabbro@ts.infn.it}
\author[Trieste,INFN]{Edoardo Milotti\corauthref{cor}}
\ead{milotti@ts.infn.it}
\corauth[cor]{Corresponding author}
\address[Verona]{Dipartimento Scientifico e Tecnologico, Facolt\`a di Scienze MM.FF.NN. \\ Universit\`a di Verona, Strada Le Grazie, 15 - CV1, 37134 Verona, Italy}
\address[INFN]{I.N.F.N. -- Sezione di Trieste}
\address[Trieste]{Dipartimento di Fisica, Universit\`a di Trieste \\ Via Valerio, 2 -- I-34127 Trieste, Italy }

\begin{abstract}
Post-transductional modifications tune the functions of proteins and regulate the collective dynamics of biochemical networks that determine how cells respond to environmental signals. For example, protein phosphorylation and nitrosylation are well-known to play a pivotal role in the intracellular transduction of activation and death signals. A protein can have multiple sites where chemical groups can reversibly attach in processes such as phosphorylation or nitrosylation. A microscopic description of these processes must take into account the intrinsic probabilistic nature of the underlying reactions. We apply combinatorial considerations to standard enzyme kinetics and in this way we extend to the dynamic regime a simplified version of the traditional models on the allosteric regulation of protein functions. We link a generic modification chain to a downstream Michaelis-Menten enzymatic reaction and we demonstrate numerically that this accounts both for thresholds and long time delays in the conversion of the substrate by the enzyme. The proposed mechanism is stable and robust and the higher the number of modification sites, the greater the stability. We show that a high number of modification sites converts a fast reaction into a slow process, and the slowing down depends on the number of sites and may span many orders of magnitude; in this way multisite modification of proteins stands out as a general mechanism that allows the transfer of information from the very short time scales of enzyme reactions (milliseconds) to the long time scale of cell response (hours).
\end{abstract}

\begin{keyword}
multisite phosphorylation \sep nitrosylation \sep threshold effect \sep biochemical model \sep network dynamics
\PACS 82.39.Fk \sep 87.16.Yc \sep 87.17.-d
\end{keyword}
\end{frontmatter}

\section{Introduction}
\label{intro}

With the advancement of biochemical techniques a huge amount of data  on the post-trasductional modification of proteins and on its role in the regulation of signal propagation through biochemical networks is now available. This knowledge is challenging our understanding of the cells' behaviour, and efforts by experts from different disciplines are now required to sort the basic dynamic principles out of experimental observations. As it has been recently pointed out \cite{mg}, this process is not obvious at all because of cultural differences that drive scientists from different disciplines either to prefer the deep molecular details of biochemical networks or to tackle the generic underlying principles. Here we follow this latter perspective, and investigate the general dynamic consequences of multiple chemical modifications of proteins within biochemical networks

Proteins can be  modified by the attachment and detachment of various chemical groups  and by means of different mechanisms. Reversible chemical modifications of proteins are now recognized to play a pivotal role in the regulation of the dynamic behaviour of biochemical networks, and therefore in the response of cells to environmental signals. The phosphorylation/dephosphorylation of  tyrosine residues, for example, allows the propagation of signals from the cell surface to the nucleus thus enabling the transcription of specific genes in response to the presence of environmental activation molecules \cite{sha,c1,c2}; another example is the nitrosylation/denitrosylation of cysteine residues as a consequence of the redox state of the intracellular environment which can activate enzymes, such as caspase-3, that partecipate in the activation of the apoptotic program that leads to controlled cell death \cite{ms}. Cell activation, proliferation and death, in turn, are at the basis of animal physiology and pathology. For example, the control of the cell cycle through protein phosphorylation is important for the activation of an immune response against foreign antigens \cite{dong}; the activation of the apoptotic program regulates tissue homeostasis and prevents the onset of autoimmune diseases and cancer \cite{hol}.

Central to most biological networks is the existence of biochemical paths that behave dynamically as on-off switches \cite{guna,qian}. The chemical modification of proteins on multiple aminoacid residues, like e.g. multisite phosphorylation, has been argued to confer a switch-like character to these proteins. One outstanding example is the retinoblastoma protein (Rb) with its 16 putative phosphorylation sites, where at least 10 of them must be phosphorylated by cyclin-dependent kinases to promote the abrupt and irreversible G1-S transition along the cell cycle \cite{ez,seville}. The switch-like behavior of proteins within biochemical networks has been traditionally modeled using the phenomenological Hill function \cite{hill}:
\begin{equation}
\label{hilleqn}
f(X)=\frac{kX^n}{1+kX^n}
\end{equation}
where $X$ is the concentration of some biochemical species, $k$ is a positive constant and $n$ is the so-called Hill coefficient. The Hill function provides switch-like sigmoidal behaviors for large $n$ values and models phenomenologically the multiple interaction of $X$ with $n$ other molecules. And yet, while the Hill function is quite good for phenomenological work it is well-known to be physically untenable for the following reasons: 1) the Hill equation implies {\it simultaneous} molecular interactions and this does not reflect a possible reaction scheme \cite{weiss}; 2) in the real world, the concentration of proteins, enzymes and substrate fluctuates randomly as a consequence of molecular diffusion processes. In addition, molecules are randomly distributed between daughter cells at mitosis and this introduces an additional stochasticity in the molecular concentrations through cellular generations \cite{bw}: the Hill function does not incorporate  the stochastic nature of the microscopic physical world and cannot be modified in this sense; 3) the sigmoidal Hill function becomes a step function, and thus describes a true threshold, only for $n {\rightarrow} {\infty}$, but experimentally,  the chemical modification of just one key protein residue can turn the function of that protein on or off. For example, each subunit of the methionine adenosyl transferase (MAT) - a member of the caspase family of cysteine proteases -  has 10 free cysteines but only cysteine 121 inhibits the activity of the enzyme when targeted by nitrosylation \cite{ms}.  The Hill function with $n=1$ reduces to a standard Michaelis-Menten function which is no longer sigmoidal and therefore does not model a threshold behavior at all. The conclusion is that the Hill phenomenology must be replaced by deeper microscopic models to obtain a better understanding of the biochemical thresholds.

From a chemical perspective, the processes leading to the chemical modification of proteins are equivalent to the allosteric regulation of enzymes whereby the substrate itself (homotropic interaction) or molecules other than the substrate (heterotropic interactions) can tune the activity of an enzyme. The allosteric theory dates back to the '60s when detailed models unifying both type of interactions where also developed, such as the famous models by Monod, Wyman and Changeaux (MWC) and by Koshland, N\'emethy and Filmer (KNF) \cite{mwc,knf}. It is not the aim of the present work to discuss these models, their differences and their adherence to experimental data, as these aspects have been fully reviewed elsewhere \cite{rubi,cb}. Some general points should nonetheless be analyzed. Both the MWC and the KNF models deal with the microscopic interactions between a chemical species and the subunits of an enzyme, the chemical species being capable of stabilizing (MWC) or inducing (KNF) a conformational change in the quaternary structure of the target subunit(s). The conformational change of the quaternary structure - but the theory has been recently extended to the conformational change of the tertiary structure of an enzyme within the very same framework of the MWC model \cite{henry} - influences in turn the enzymatic activity. In other words, these models focus on the mechanics  (i.e. modifications of the protein structure) and not on the dynamics of the allosteric effect. And in fact, the models rely on a set of assumptions whose common one is the steady state hypothesis: protein complexes are in equilibrium with a very large number of ligands whose concentration is always constant \cite{mwc,knf,rubi,cb}. This assumption allows the authors to linearize the model equations which, however, do still contain a number of parameters whose values must be estimated by fitting equations to experimental data. It has been pointed out that the high number of adjustable parameters hampers the validation of a given allosteric model \cite{rubi} and, we add, also hinders the inclusion of such equations into large biochemical ensembles. Simpler models that do take into account the real variations of the ligand concentrations within a cell are therefore required to describe the dynamic behaviour of protein networks. Of course a model of this kind must be based on a simplification of the molecular details underlying the allosteric effect. We propose here to focus the analysis on the dynamic properties of the allosteric effect and we put forward a simple and very general microscopic model of the multiple chemical modifications of proteins. Our analysis does not explain how the allosteric effect is realized structurally by a given protein but it shows how certain important properties of biochemical networks - such as thresholds, long delays and stability - emerge from a simplified analysis of allosteric dynamics.  

\section{Equilibrium concentrations of the modified protein forms}
\label{eqval}

\begin{figure}
\includegraphics[width= 14cm]{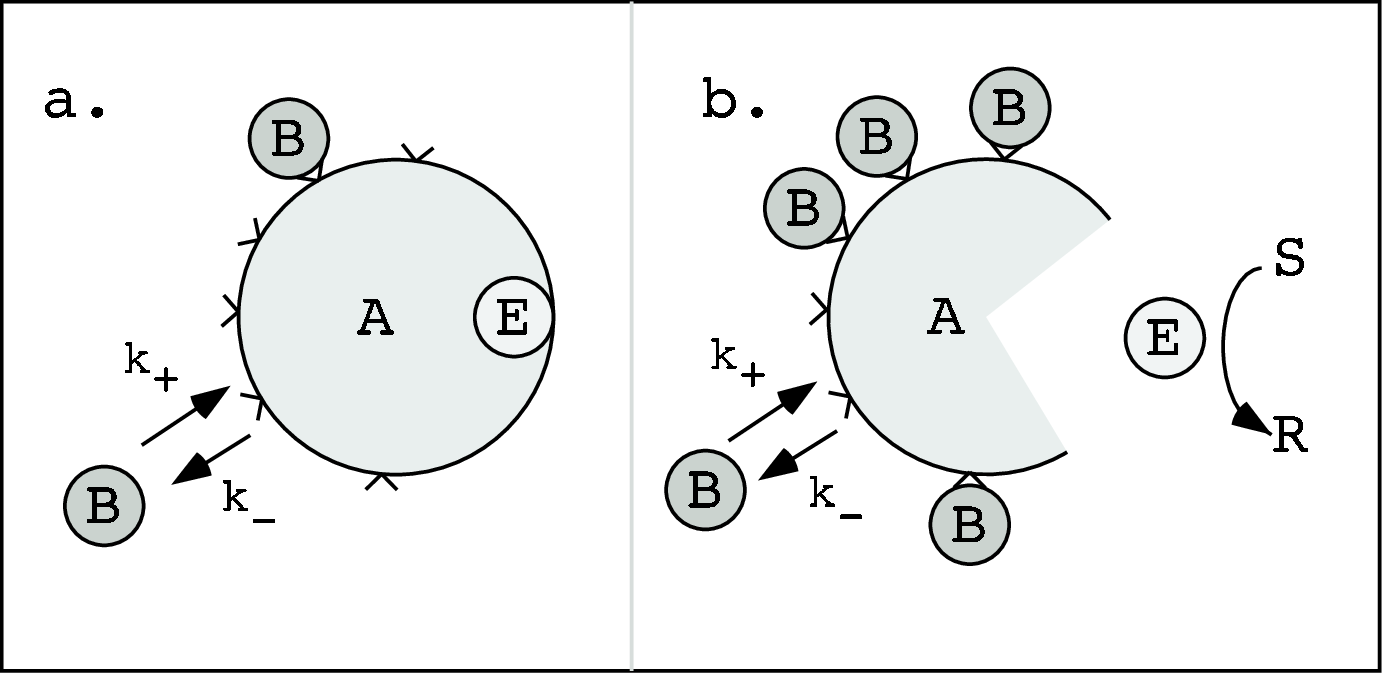}
\caption{\label{fig1} {\it a.} Molecule A has $N$ phosphorylation sites: here the process is schematically represented by molecules B that react with each site with on-off rates  $k_+$, $k_-$. {\it b.} When at least $n$ sites out of the possible $N$ sites are phosphorylated, molecule A releases an enzyme E which catalyses a Michaelis-Menten reaction that converts a substrate S into a product R.}
\end{figure}

We start by considering the reaction scheme shown in figure 1, which is basically equivalent to that considered by Monod et al. \cite{mwc},  where a chemical species B can modify a number of sites on protein A. We also assume that the  sites are all equivalent and that the modification dynamics for each site is independent from those of the other sites: this means that we consider the states $A_n$ and the transition chain shown in figure 1. If the single chemical modification dynamics is fast with respect to the observation time we can forget the dynamics of the transition chain and concentrate instead on the equilibrium probabilities. Indeed, if $p$ is the probability that any given site is chemically modified, the equilibrium probability $P_n$  that at least $n$ molecules of the species B bind to A is given by the sum of the corresponding probabilities from a binomial distribution and therefore if the volume of reaction $V$ contains $\nu_A = N_A V \left[ A \right]$ molecules of type A, where $N_A$ is the Avogadro constant and the square brackets denote molar concentrations, the average number of A molecules with at least $n$ occupied sites is:
\begin{equation}
\label{prob2}
\nu_A P_n = \nu_A \sum_{l=n}^{N} \binom{N}{l} p^l (1-p)^{N-l}
\end{equation}
Every single modification reaction can be represented as a generic bimolecular reaction, $A+B \underset{k_-}{\overset{k_-}{\leftrightarrows}} AB$, so that the following equations hold
\begin{eqnarray}
\label{diffsys}
\nonumber
&& \frac{d\left[ A \right]}{dt} = - k_ +  \left[ A \right]\left[ B \right] + k_ -  \left[ AB \right] \\
&& \frac{d\left[ B \right]}{dt} = - k_ +  \left[ A \right]\left[ B \right] + k_ -  \left[ AB \right] \\
\nonumber
&& \frac{d\left[ AB \right]}{dt} = k_ +  \left[ A \right]\left[ B \right] - k_ -  \left[ AB \right]
\end{eqnarray}
The reaction dynamics (\ref{diffsys}) leads to the conservation equations $[A] +[AB]=[A]_0$ and $[B] +[AB] =[B]_0$, where $[A]_0$ and $[B]_0$ are the initial concentrations, while at equilibrium the derivatives vanish and the condition $\left[ AB \right]  = (k_+/k_-)[A][B]$ holds; combining these conditions we find an equation that can easily be solved, and we obtain 
\begin{eqnarray}
\label{occ1}
\nonumber
[AB] &= & \frac{1}{2} \left\{ \left( [A]_0  +[B]_0  + \frac{k_-}{k_+} \right) \right. \left. \right. \\
&& \left.- \sqrt {( [A]_0 -[B]_0)^2  + 2( [A]_0 +[B]_0) \frac{k_-}{k_+} + \left(\frac{k_-}{k_+} \right)^2 }  \right\}
\end{eqnarray}
Referring to figure \ref{fig1} we reinterpret the result (\ref{occ1}) as a concentration of occupied sites, so that the probability $p$ is
\begin{eqnarray}
\label{sitefraction}
p & = &\frac{\left[ \text{occupied sites} \right]} {N[A]_0 }  \\
\nonumber
& = & \frac{1}{2N[A]_0}  \left\{ 
\left( N[ A]_0  +[ B]_0  + \frac{k_-}{k_+} \right) \right. \\
&& \left. - \sqrt{( N[ A]_0  -[ B]_0)^2  + 2( N[ A]_0  +[ B]_0)\frac{k_-}{k_+} +\left( \frac{k_-}{k_+}\right)^2 } 
\right\} 
\end{eqnarray}
where we note in passing that the dissociation constant $k_-/k_+$ depends on temperature (in general $k_-/k_+ = \exp\left(\Delta G/RT \right)$, and $\Delta G$ is the Gibbs free energy.

\section{Threshold and threshold robustness}
\label{thrsec} 

Now we assume that the molecule A somehow switches to an activated form A$^*$ when at least $n$ modified sites are occupied (see figure 1), and that the process is reversible. This might occur because of a conformational switch \cite{choi} or because A releases some reaction product, but the detailed description of this step is not important, what really matters is that a change occurs only if at least $n$ sites are involved. It is worth noting that this analysis includes the case $n=1$. Only few molecules are actually activated by this mechanism when $p$ is small. For example, in a spherical cell with a radius of 5 $\mu$m and a corresponding volume $\approx 5 \cdot 10^{-16} \mathrm{m}^3$, where proteins have concentrations below 10 $\mu$M, there are fewer than $3\cdot 10^3$ molecules of each kind of protein, and even with the not so small value $p \approx 0.1$ we find, e.g., $P_n \approx 4.5 \cdot 10^{-7}$  for $N = 16$ and $n = 10$ (as for the Rb protein), so that in this case there is on average much less than 1 activated molecule. This means that molecular discreteness is important, and sets a natural threshold for the process set in motion by the activated form of A: the process cannot proceed at all if, on average, there is less than 1 activated molecule and the threshold is given by the probability $p$ (and by the corresponding concentrations in equation (\ref{sitefraction})) that solves the equation $\nu_A P_n =1$. Since $p < 1$, equation (\ref{prob2}) can be approximated by the simple power law
\begin{equation}
\label{plaw}
\nu_A P_n \approx  N_A V [ A]_0  \binom{N}{n} p^n
\end{equation}
and this means that for $n = 10$ as in the example, the threshold is very sharply defined (see figure \ref{fig3}).
The threshold defined by the solution of the equation $\nu_A P_n =1$ , with $n > 1$, is not only sharp but very robust as well. There are many sources of potentially destructive fluctuations, like the variable number of A molecules that could change as a consequence of the unequal distribution of enzymes and substrates between daughter cells at mitosis \cite{bw}, but it is easy to see that these fluctuations are damped down, and that the damping-off is more successful for large $n$'s. 
When the $k_-/k_+$ ratio is negligible with respect to $[A]_0$ and $[B]_0$, the probability (\ref{sitefraction}) can be approximated as follows:
\begin{equation}
p([B]_0) \approx \frac{1}{2N}\left\{ {\left( {N + \frac{[B]_0 }{[A]_0}} \right)
 - \left| {N - \frac{[B]_0}{[A]_0}} \right|} \right\} = 
\left\{ {\begin{array}{ccc}
[B]_0/(N [A]_0) & \mathrm{if} & N >[B]_0/[A]_0  \\
1&\mathrm{if} & N \leq [B]_0/[A]_0  \\
\end{array} } \right.
\end{equation}
with a breakpoint at $[B]_0 = N [A]_0$. In general, for low concentration $[B]_0$, expression (\ref{sitefraction}) depends linearly on $[B]_0$:
\begin{equation}
\label{approxPn}
p([B]_0) \approx \frac{[B]_0}{N [A]_0 + (k_-/k_+)}
\end{equation}
Using the approximate expression (\ref{approxPn}) and equation (\ref{plaw}) we find the threshold value: 
\begin{equation}
[B]_{0,thr} = \left( N [A]_0 + \frac{k_-}{k_+} \right) \left\{ N_A V [A]_0 \binom{N}{n}  \right\}^{-1/n}
\end{equation}
If we use standard formulas for the analysis of random fluctuations of experimental quantities, we find
\begin{equation}
\frac{\operatorname{var} [B]_{0,thr}}{([B]_{0,thr})^2} =\frac{1}{n^2} \frac{\operatorname{var}{V}}{V^2} + \left( \frac{N[A]_0(n-1)-(k_-/k_+)}{n(N[A]_0+(k_-/k_+))} \right)^2 \frac{\operatorname{var} [A]_{0}}{([A]_{0})^2}
\end{equation}
and the end result is that volume fluctuations are damped by a factor $n$, so that, e.g., a 10\% volume fluctuation produces a mere 1\% change in the threshold position for $n= 10$, while the damping of substrate fluctuations varies from slightly sublinear (so that a 10\% fluctuation of $[A]_0$ produces a 9\% change in the threshold position for $n=10$) to a strong damping condition like in the case of volume (for $k_-/k_+ \gg N[A]_0$) . 

\section{Downstream Michaelis-Menten step}
\label{MandM} 

As shown in figure \ref{fig1}, molecules A activated by multiple chemical modifications are supposed to release an enzyme E that catalyses a Michaelis-Menten reaction where a substrate S is converted into a product R. The downstream reaction catalyzed by E writes:
\[
E+S \underset{k_2}{\overset{k_1}{\leftrightarrows}} ES \stackrel{k_3}{\longrightarrow} E + R
\]
and we use the Michaelis-Menten equations with the usual quasi-steady state assumption (QSSA, see, e.g. ref. \cite{schnell} and references therein), so that the substrate consumption is described by the single differential equation
\begin{equation}
\label{MM}
\frac{{d\left[ S \right]}}
{{dt}} \approx  - \frac{{k_3 k_1 \left[ E \right]_0 \left[ S \right]}}
{{k_1 \left[ S \right] + \left( {k_2  + k_3 } \right)}} =  - \frac{{k_3 \left[ E \right]_0 \left[ S \right]}}
{{K_m  + \left[ S \right]}}
\end{equation}
where $K_m = (k_2+k_3)/k_1$ and the production of R is related to the consumption rate by the equality
$d[ R]/dt = - d[ S]/dt$
From the considerations above, it is easy to see that the concentration of the enzyme E equals the concentration of  the molecules A with at least $n$ occupied sites:
\begin{equation}
\label{enzyme}
[E]_0 = [A]_0 \sum_{l=n}^{N} \binom{N}{l} p^l (1-p)^{N-l}
\end{equation}
then, if we combine equations (\ref{MM}) and (\ref{enzyme}), we can integrate numerically equation (\ref{MM}). Figure \ref{fig4} shows the behaviour of [R] obtained with parameters in a range common to many biochemical reactions in the cell. Here we took the concentration $[B]_0$  to be a linear time-dependent quantity: $[B]_0 = 16\cdot (10 \mu\mathrm{M}) \cdot(10^{-5} \mathrm{s}^{-1}) t$, so that $[B]_0$ reaches the critical value $N [A]_0$ at $t = 10^5$ s. Figure \ref{fig5} shows that as the chemical modification level $n$ grows the time delay  $\Delta t$ (time taken to reach 50\% of the maximum product concentration) spans several orders of magnitude (from a few seconds up to a few hours). The production of R is not just delayed, it is also slowed down: figure \ref{fig5} shows that the production interval spans several orders of magnitude as well. 

\section{Conclusion}
\label{conc}

In the mathematical modeling of cooperative reaction kinetics, threshold effects are commonly achieved with phenomenological Hill equations, however here we no longer require this kind of phenomenological modeling and both thresholds and  time delays have a clear meaning. This is particularly important in models of  the behavior of complex biochemical systems, as time delays coupled to negative feedback regulatory circuits are required to produce bistability and limit cycle behavior  \cite{yang}. We have neglected the coupling between the modification step and the downhill Michaelis-Menten step, because the probabilistic model does not describe an actual reaction dynamics. The Michaelis-Menten step assumes that part of the enzyme is bound with the substrate to form the complex ES, and this acts as a store of E, which is released with a characteristic relaxation time  $\tau = \left[ k_1[S] + (k_2+k_3)\right]^{-1}$ \cite{cb}. For this reason we expect the downhill reaction to damp off any noise in the $[B]_0$ concentration signal: thus the combination of the modification step with the downhill (irreversible) Michaelis-Menten step, makes the whole system even more robust, and resistant to environmental changes. 

All this is particularly intriguing from an evolutionary point of view: proteins with a high number of residues that can be modified chemically might in fact have been selected because this would intrinsically stabilize the threshold effect and hence the overall dynamics of the biochemical network operating in highly perturbed environments.



\pagebreak


\begin{figure}
\includegraphics[width= 14cm]{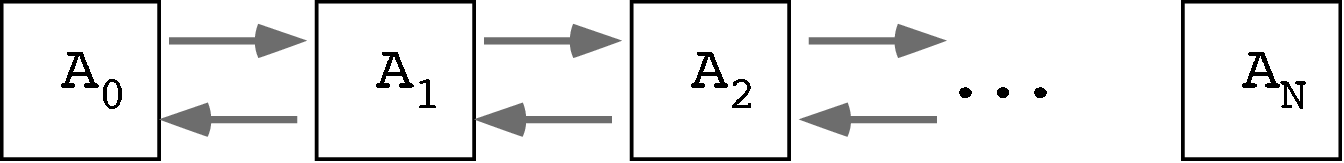}
\caption{\label{fig2} Schematic representation of the phosphorylation-dephosphorylation chain. We assume that all sites are equivalent and that the enzyme state depends only on the number of phosphorylated sites, so that there are only $N+1$ different states which are labeled $A_l$. The arrows represent the phosphorylating-dephosphorylating transitions.}
\end{figure}

\begin{figure}
\includegraphics[width= 14cm]{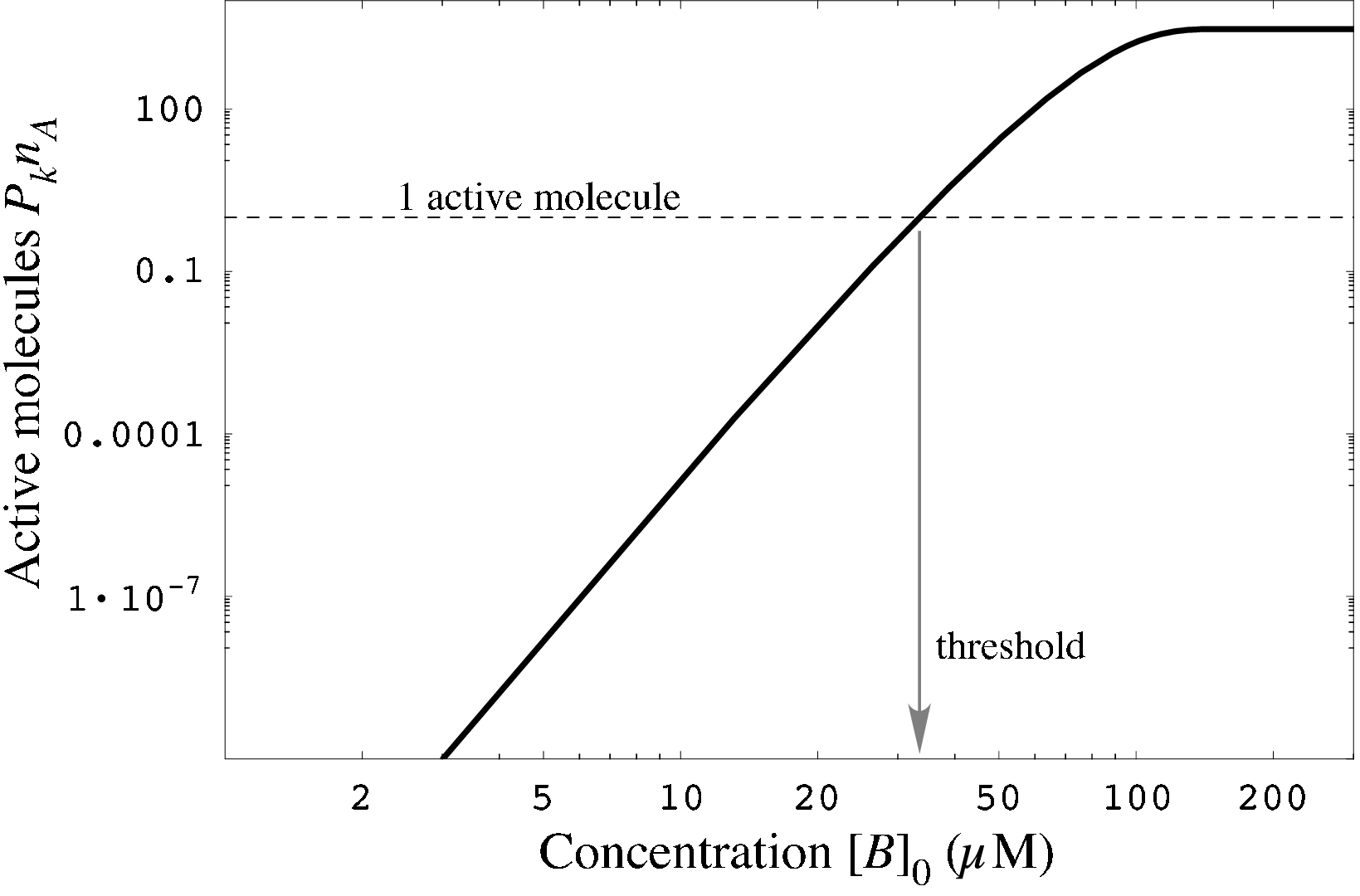}
\caption{\label{fig3} Log-log plot of the (average) number of active A$^*$ molecules vs. $[B]$, obtained from equations (\ref{prob2}) and (\ref{sitefraction}). The number of active A$^*$ molecules with at least $n = 10$ phosphorylated sites over $N = 16$ possible sites has been calculated using equation (\ref{prob2}) with the following parameters: $[A]_0 = 10 \mu$M; $k_-/k_+ = 10^{-6}$ M (this is in the observed range for the dissociation constant of ATP with cyclin-dependent kinases \cite{clare}) and with a cell volume $\approx 5\cdot 10^{-16} \mathrm{m}^3$, so that the cell contains approximately 3000 A molecules as in the example discussed in the text. The dashed line marks the threshold level $\nu_A P_n =1$.}
\end{figure}

\begin{figure}
\includegraphics[width= 14cm]{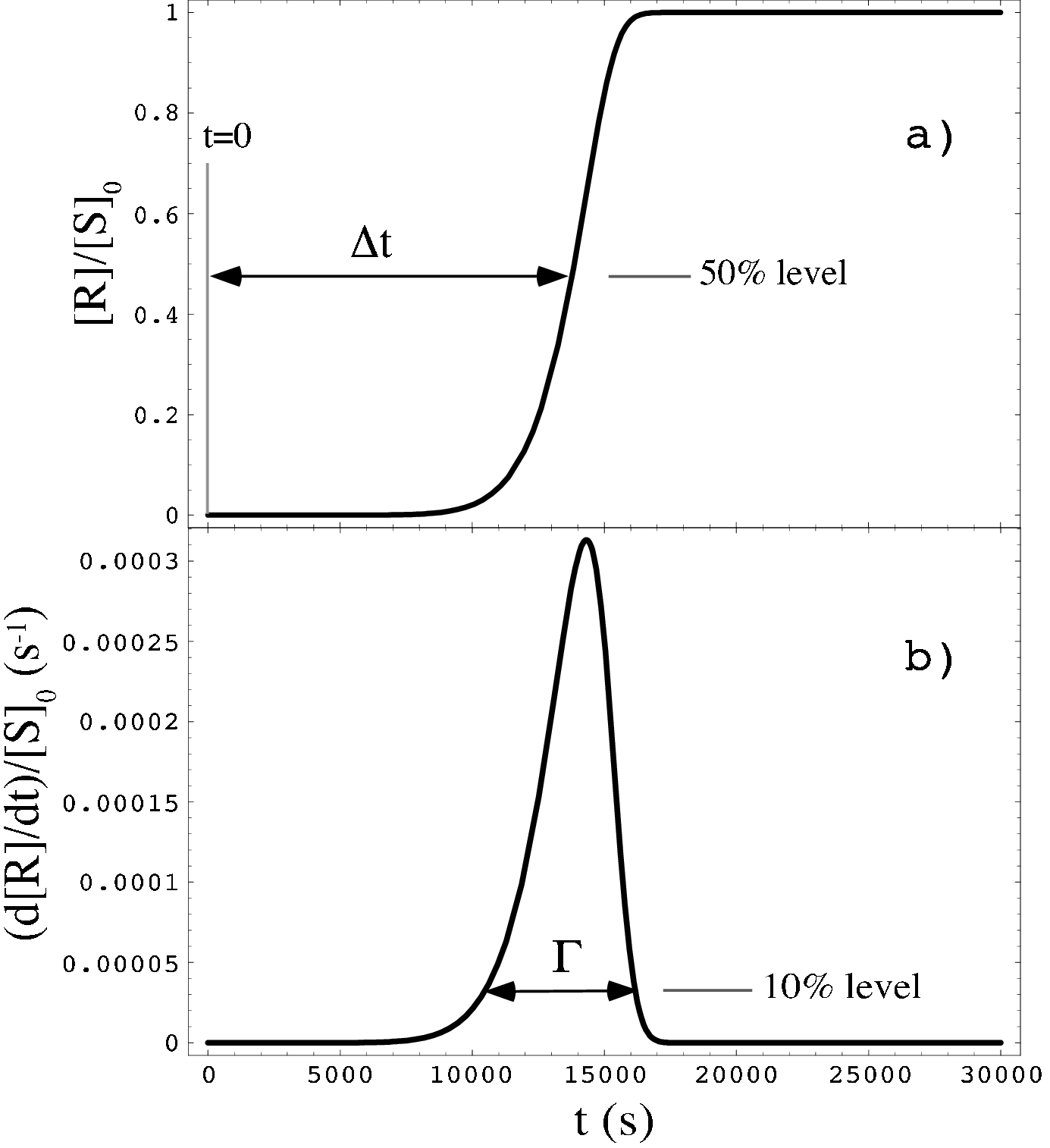}
\caption{\label{fig4} a) Relative concentration $[R]/[S]_0$ obtained from the numerical integration of equation (\ref{MM}) with condition (\ref{enzyme}); in this case  $K_m = 1$mM, $k_3  = 10^4 \mathrm{s}^{-1}$, $[A]_0 =10 \mu$M, $[S]_0 = 1 $mM, $k_-/k_+ = 10^{-6}$ M, $N = 16$, $n = 10$, and with a linear growth law $[B]_0 = 16á(10 \mu\mathrm{M})\cdot (10^{-5} \mathrm{s}^{-1}) t$: in this way the concentration of $[B]_0$ equals the critical concentration $N [A]_0$ at $t = 10^5$s. $\Delta t$ is the time delay needed to reach 50\% of the maximum product concentration $[R]_{max} = [S]_0$.
b) Scaled derivative $\frac{1}{[S]_0}\left( \frac{d[R]}{dt} \right)$: the width $\Gamma$ (measured at 10\% of the peak value) gives an estimate of the process duration.}
\end{figure}

\begin{figure}
\includegraphics[width=14cm]{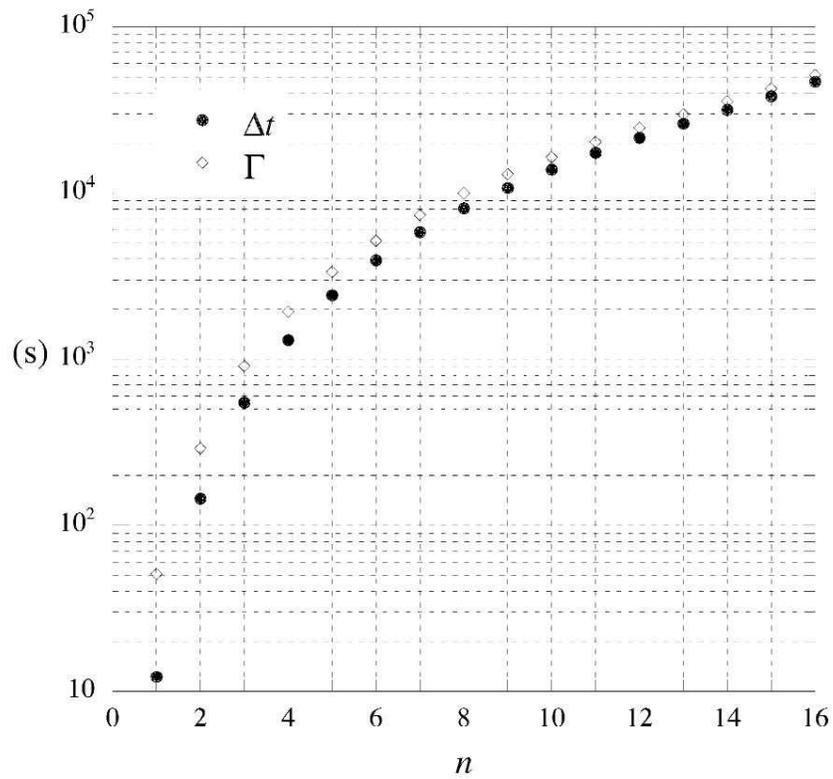}
\caption{\label{fig5} Plot of the time delay $\Delta t$  and of the width $\Gamma$ vs. the number of phosphorylated sites $n$ from the integration in figure \ref{fig4}. As $n$ grows both $\Delta t$ and $\Gamma$  span several orders of magnitude, and range from a few seconds to several hours.}
\end{figure}

\end{document}